\documentstyle[preprint,aps]{revtex}
\tighten
\draft
\widetext
\input epsf
\preprint{HUTP-98/A073, NUB 3189}
\bigskip
\bigskip
\begin{document}
\title{Novel Extension of MSSM and ``TeV Scale'' Coupling Unification}
\medskip
\author{Zurab Kakushadze\footnote{E-mail: 
zurab@string.harvard.edu}}
\bigskip\bigskip
\address{
Lyman Laboratory of Physics, Harvard University, Cambridge, MA 02138\\
and\\
Department of Physics, Northeastern University, Boston, MA 02115}
\date{November 22, 1998}
\bigskip
\medskip
\maketitle

\begin{abstract}
{}Motivated by the coupling unification problem, we propose a novel extension of the
Minimal Supersymmetric Standard Model. One of the predictions of this extension is existence of new states neutral under $SU(3)_c\otimes SU(2)_w$ but charged under $U(1)_Y$. The mass scale for these new states can be around the mass scale of the electroweak Higgs doublets. This suggests a possibility of their detection in the present or near future collider
experiments. Unification of gauge couplings in this extension is as precise (at one loop)
as in the MSSM, and can occur in the TeV range.  

\end{abstract}

\section{Introduction}

{}Supersymmetry provides an elegant solution to the naturalness and hierarchy problems. The Minimal Supersymmetric Standard Model (MSSM) is a viable candidate for describing physics
above the electroweak scale. At present there is no direct evidence for supersymmetry. Nonetheless, unification of gauge couplings \cite{gut} in the MSSM provides an indirect hint
for new physics at LHC.

{}The gauge coupling unification in the MSSM occurs with a remarkable precision. It is, however, reasonable to ask whether the unification scale is indeed $M_{\small{GUT}}\approx 2\times 10^{16}~{\mbox{GeV}}$ as 
predicted via extrapolating the LEP data by assuming 
the MSSM matter content all the way up to $M_{\small{GUT}}$ (that is, by assuming the standard 
``desert'' scenario).   
In particular, one might wonder whether it is possible to consider an extension of the MSSM
where unification of couplings occurs with just as good precision as in the MSSM but, say,
at a lower scale. In fact, one could also require that such an extension {\em explain} why
couplings unify in the MSSM at all, that is, why this unification is {\em not} just an ``accident''. 

{}The question of where and how the gauge coupling unification occurs is certainly important
from the phenomenological viewpoint. There is, however, additional motivation to address this issue. In string theory, which is the only known theory that consistently incorporates quantum gravity, the gauge and gravitational couplings generically are expected to unify 
(up to an order one factor due to various thresholds) at the string scale
$M_s=1/\sqrt{\alpha^\prime}$. {\em A priori} the string scale can be anywhere between the electroweak 
scale $M_{ew}$ and the Planck scale $M_P=1/\sqrt{G_N}$ (where $G_N$ is
the Newton's constant). Thus, in the ``Brane World'' picture \cite{BW} the Standard Model fields
reside inside a $p$-brane, whereas gravity lives in a larger (10 or 11) dimensional bulk of the space-time. For instance, if we assume that the bulk is ten dimensional,  then the four dimensional gauge and gravitational couplings 
scale as\footnote{For illustrative purposes here we are using the corresponding tree-level relations in Type I (or Type I$^\prime$) theory.} $\alpha\sim g_s/V_{p-3} M_s^{p-3}$ respectively $G_N\sim g_s^2/V_{p-3} V_{9-p} M_s^8$, where $g_s$ is the string coupling, and $V_{p-3}$ and $V_{9-p}$ are the compactification volumes inside and transverse to the $p$-brane, respectively. For $3<p<9$ there are two {\em a priori} independent volume factors, and, for the fixed gauge coupling $\alpha$ (at the
unification, that is, string scale) and four dimensional Planck scale $M_P$, the string scale is
not determined \cite{witt}.

{}In fact, in the brane world picture {\em a priori} the string scale can be as low as desired as long as it does not directly contradict current experimental data. In \cite{lyk} a possibility of
having $M_s$ as low as TeV was discussed. In \cite{TeV} it was proposed that $M_s$ as well as the fundamental (10 or 11 dimensional) Planck scale can be around TeV. The observed weakness of the four dimensional gravitational coupling then requires presence of at least two large ($\gg 1/M_s$) compact directions (which are transverse to the branes on which the Standard Model fields are localized). A general discussion of possible brane world embeddings
of such a scenario was given in \cite{anto,ST}. 
In \cite{TeVphen} various phenomenological 
issues were discussed in the context of the TeV string scenario, and it was argued that this possibility does not appear to be automatically ruled out. However, in such a scenario, as well as in any scenario where $M_s\ll
M_{\small{GUT}}$, the gauge coupling unification at $M_s$ would have to occur in a way
drastically different from the usual MSSM unification. In the brane world picture, however, there
appears to exist a mechanism \cite{dien} for lowering the unification scale. Thus, let the ``size'' $R$ of the compact dimensions inside of the $p$-brane (where $p>3$) be somewhat large compared with $1/M_s$. Then the evolution of the gauge couplings above the Kaluza-Klein (KK) 
threshold $1/R$ is no longer logarithmic but power-like \cite{TV}. (This can alternatively be thought of as considering a higher dimensional theory at energy scales larger than $1/R$.) This observation was used in \cite{dien} to argue that the gauge coupling unification might occur at a scale
(which would be identified with the string scale) much lower than $M_{\small{GUT}}$. The question that needs to be addressed in this context, however, is whether there exists an appropriate extension of the MSSM where the gauge coupling unification can indeed occur
via such a mechanism. (For other recent works on TeV scale string/gravity
scenarios, see, {\em e.g.}, \cite{related}. For other scenarios with
lowered string scale and related works, see, {\em e.g.},
\cite{other}. TeV scale compactifications were studied in \cite{quiros} in
the context of supersymmetry breaking.)

{}Motivated by the above considerations, in this paper we propose a novel extension of the MSSM. In this extension the (one-loop) 
gauge coupling unification occurs with just as good precision as
in the MSSM. The unification scale, which we will refer to as $M_s$, 
can, however, be much lower than $M_{\small{GUT}}$. The light modes in our model consist of the MSSM fields plus two new ${\cal N}=1$ chiral superfields $F_{\pm}$ transforming in 
$({\bf 1},{\bf 1})(\pm 2)$ of $SU(3)_c\otimes SU(2)_w\otimes U(1)_Y$ (the hypercharge is given
in parentheses). The massive KK modes (corresponding to $p-3>0$ compact
directions arising in the embedding of this model in a $p$-brane) have
${\cal N}=2$ supersymmetry from the four dimensional viewpoint. (We will
give the precise massive KK 
spectrum in section II.)
An interesting prediction of the extension we propose here is possible 
existence of new light modes
$F_\pm$. Since these states are ``vector-like'' (just as the MSSM electroweak Higgs doublets
$H_\pm$), it is possible that their mass scale is around that of the Higgs doublets.
(In particular, the ``$\mu^\prime$-term'' $\mu^\prime F_+ F_-$ for the $F_\pm$ fields can be comparable to the $\mu$-term $\mu H_+ H_-$ for the Higgs doublets.) This suggests a possibility that, if the fields $F_\pm$ indeed exist, they could be detected in the present or near future collider experiments. 
  
{}The rest of this paper is organized as follows. In section II we give the precise definition
for the extension of the MSSM we propose here.
In section III we discuss the gauge coupling unification in our models. In section IV we discuss  possible brane world embeddings with some properties of the KK spectrum arising in our model. In section V we summarize our results and discuss some open questions.   
Some details of KK threshold computation are relegated to appendix A.

\section{The Model}

{}In this section we describe in more detail the extension of the MSSM outlined in Introduction.
For later convenience, let us fix the notation for the MSSM fields. We have ${\cal N}=1$ supersymmetric $SU(3)_c\otimes SU(2)_w \otimes U(1)_Y$ gauge theory with the following chiral supermultiplets:
\begin{eqnarray}
 && Q_i=3\times ({\bf 3},{\bf 2})(+1/3)~,~~~
 D_i=3\times ({\overline {\bf 3}},{\bf 1})(+2/3)~,~~~
 U_i=3\times ({\overline {\bf 3}},{\bf 1})(-4/3)~,\nonumber\\
 && L_i=3\times ({\bf 1},{\bf 2})(-1)~,~~~E_i=3\times ({\bf 1},{\bf 1})(+2)~,\nonumber\\
 && H_+=({\bf 1},{\bf 2})(+1)~,~~~H_-=({\bf 1},{\bf 2})(-1)~.\nonumber
\end{eqnarray}
Here the $SU(3)_c\otimes SU(2)_w$ quantum numbers are given in bold font, whereas the $U(1)_Y$ hypercharge is given in parentheses. The three generations $(i=1,2,3)$ of quarks 
and leptons are given by $Q_i,D_i,U_i$ respectively $L_i,E_i$, whereas $H_\pm$ correspond to the electroweak Higgs doublets. Next, consider the extension of the MSSM obtained via adding two new chiral superfields $F_\pm$:
\begin{eqnarray}
 F_+=({\bf 1},{\bf 1})(+2)~,~~~F_-=({\bf 1},{\bf 1})(-2)~.\nonumber
\end{eqnarray}
In the following we will use the collective notation $\Phi\equiv (V,Q_i,U_i,D_i,L_i,E_i,H_\pm,
F_\pm)$, where $V$ stands for the ${\cal N}=1$ vector superfields transforming in the adjoint
of $SU(3)_c\otimes SU(2)_w \otimes U(1)_Y$. (Here we should point out that
incorporating right-handed neutrinos in this model is straightforward in
the sense that they do not affect the unification prediction as they are
neutral under $SU(3)_c \otimes SU(2)_w \otimes U(1)_Y$.)  

{}We now wish to consider a straightforward lifting of the above model with ${\cal N}=1$
superfields $\Phi$ to ${\cal N}=2$ supersymmetry. In this process the ${\cal N}=1$ vector
multiplets $V$ are promoted to ${\cal N}=2$ vector multiplets ${\widetilde V}=V\oplus\chi$,
where $\chi$ stands for the ${\cal N}=1$ chiral superfields transforming in the adjoint of $SU(3)_c\otimes SU(2)_w \otimes U(1)_Y$. Similarly, all of the ${\cal N}=1$ chiral supermultiplets are promoted to the corresponding ${\cal N}=2$ hypermultiplets which differ from the former by the additional ${\cal N}=1$ chiral supermultiplets of opposite chirality but the same gauge quantum numbers. We will refer to these additional ${\cal N}=1$ chiral superfields
as $Q^\prime_i,U^\prime_i,D^\prime_i,L^\prime_i,E^\prime_i,H^\prime_\pm,F^\prime_\pm$.
Let ${\widetilde \Phi}$ be the collective notation for the
${\cal N}=2$ superfields. Then we can write ${\widetilde \Phi}=\Phi\oplus\Phi^\prime$, where
$\Phi^\prime\equiv (\chi,Q^\prime_i,U^\prime_i,D^\prime_i,L^\prime_i,E^\prime_i,H^\prime_\pm,F^\prime_\pm)$. 

{}We can view the four dimensional ${\cal N}=2$ supersymmetric model with the superfields
${\widetilde \Phi}$ as the low energy limit of, say, the corresponding six (five) dimensional ${\cal N}=1$ supersymmetric theory compactified on a two-torus $T^2$ (circle $S^1$). 
More precisely, let us consider a gauge theory with eight supercharges living in the world-volume of some number
of coincident D$p$-branes \cite{polchi} (where $p=4$ or 5). For the sake of concreteness let us focus on the case of D5-branes (as the discussion for D4-branes is similar). Let two of the spatial dimensions inside of the D5-branes be compactified on $T^2$, which, for the sake of simplicity, we will take to be a product of two identical circles $T^2=S^1\otimes S^1$ with the
radius $R$. Then the effective field theory below the KK threshold $1/R$ is the four dimensional ${\cal N}=2$ supersymmetric gauge theory. The masses of the KK modes are given by
\begin{equation}
 M^2_{\bf m} ={1\over R^2} {\bf m}^2~,
\end{equation} 
where the vector ${\bf m}=(m_1,m_2)$ ($m_1,m_2\in {\bf Z}$) labels the KK levels. The quantum numbers of the fields at each KK level are given by ${\widetilde \Phi}$. Note that both
the light (that is, ${\bf m}=(0,0)$) and massive KK modes have ${\cal N}=2$ supersymmetry from the four dimensional viewpoint. 

{}Next, we would like to orbifold this ${\cal N}=2$ KK theory to obtain a model with ${\cal N}=1$
supersymmetric light spectrum. That is, the orbifold group $\Gamma$ must be a finite discrete subgroup of the $SU(2)_R$ global 
R-parity group of the ${\cal N}=2$ superalgebra. More precisely, 
$\Gamma$ must be a subgroup of a $U(1)_R$ subgroup of $SU(2)_R$ since we wish to preserve ${\cal N}=1$ supersymmetry. This implies that ${\Gamma}$ must be isomorphic to ${\bf Z}_N$. It is straightforward to consider the most general case, but for the sake of simplicity we will focus
on the simplest possibility, that is, ${\Gamma}\approx {\bf Z}_2$. The action of the generator $g$ of this ${\bf Z}_2$ breaks $SU(2)_R$ down to $U(1)_R$. In particular, 
under the action of $g$ the fields $\Phi$ are even, whereas the fields $\Phi^\prime$ are odd.
Here we are assuming that $g$ acts trivially on the $SU(3)_c\otimes SU(2)_w\otimes U(1)_Y$
gauge quantum numbers as well as on the ``flavor'' quantum numbers (corresponding to the three generations of quarks and leptons). 
We also need to specify the action of $g$ on the KK quantum numbers
labeled by ${\bf m}$.

{}Thus, let $x,y$ be the real coordinates corresponding to the two directions
compactified on $S^1\otimes S^1$. We can then take the action of $\Gamma$ on $x,y$ to be
given by $g:(x,y)\rightarrow (-x,-y)$. This implies that $g:(m_1,m_2)\rightarrow (-m_1,-m_2)$. Then for a given ${\bf m}\not=(0,0)$ the linear combination $(|{\bf m}\rangle+
|-{\bf m}\rangle)/\sqrt{2}$ is even under the action of $g$, whereas $(|{\bf m}\rangle-
|-{\bf m}\rangle)/\sqrt{2}$ is odd. Also, the state $|{\bf m}\rangle = |(0,0)\rangle$ is even under ${\bf Z}_2$. The spectrum of the orbifold model is obtained by keeping the states even under $g$, while projecting out the states that are odd. This gives us the following spectrum:
\begin{eqnarray}
 &&|(0,0)\rangle\otimes |\Phi\rangle~,\nonumber\\
 &&{1\over \sqrt{2}} (|(m_1,m_2)\rangle+|(-m_1,-m_2)\rangle)\otimes 
 |\Phi\rangle~,\nonumber\\
 &&{1\over \sqrt{2}} (|(m_1,m_2)\rangle-|(-m_1,-m_2)\rangle)\otimes 
 |\Phi^\prime\rangle~,\nonumber
\end{eqnarray}
where without loss of generality we can assume that $m_1>0$. Thus, the light modes
are given by the ${\cal N}=1$ superfields $\Phi$, whereas the massive KK levels are populated by ${\cal N}=2$ superfields ${\widetilde \Phi}$. (Note that at massive KK levels we can combine the states with $\Phi$ and $\Phi^\prime$ quantum numbers into ${\cal N}=2$ superfields with ${\widetilde \Phi}$ quantum numbers.)

\subsection{Refinements}

{}In the above construction we have assumed that the action of the orbifold group $\Gamma$ on the ``flavor'' quantum numbers is trivial. Here we can consider the following generalization. First, let us modify the parent ${\cal N}=2$ theory in the following way. Let each KK level be populated by the superfields ${\widetilde \Phi}=\Phi\oplus
\Phi^\prime$, where now we take $\Phi=(V,Q_r,U_r,D_r,L_r,E_r,H_\pm,F_\pm)$, and
$\Phi^\prime=(\chi,Q^\prime_r,U^\prime_r,D^\prime_r,L^\prime_r,E^\prime_r,H^\prime_\pm,
F^\prime_\pm)$. Here $r=1,\dots,n_f$, where previously we had $n_f=3$, but now we will take $n_f$ to be arbitrary. 
If we now perform the ${\bf Z}_2$ orbifold with the orbifold action on the gauge and ``flavor'' quantum numbers being trivial, the light spectrum of the corresponding ${\cal N}=1$ model will have $n_f$ chiral families of quarks and leptons.   

{}To obtain a low energy spectrum with three chiral families we can proceed as follows. 
First, let us discuss the cases where $n_f>3$. We can modify the ${\bf Z}_2$ action so that the generator of ${\bf Z}_2$ acts non-trivially on the ``flavor'' quantum numbers. In particular, let us assume that $n^+_f$ ``flavors'' are ${\bf Z}_2$ even, whereas $n^-_f$ ``flavors'' are ${\bf Z}_2$ odd, where $n^+_f+n^-_f=n_f$, and $n^+_f-n^-_f=3$.
Then it is not difficult to see that we will have the low energy spectrum consisting of ${\cal N}=1$ $SU(3)_c\otimes SU(2)_w\otimes U(1)_Y$ gauge theory with the matter given by the chiral superfields $H_\pm,F_\pm$ as well as $n^+_f$ left-handed
and $n^-_f$ right-handed chiral generations of quarks and leptons. Thus, we have three chiral families plus $n^-_f$ ``vector-like'' families. (The latter can generically acquire masses of order of $M_{ew}$ or higher, so that their presence need not contradict the current experimental data.) The massive KK spectrum is ${\cal N}=2$ supersymmetric as before. The only difference is that the number of hypermultiplets which are the ${\cal N}=2$ counterparts of the MSSM chiral generations is now $n_f$. 

{}Next, let us consider the cases where $n_f<3$. Here the analysis is the same as above except
that now $n^+_f-n^-_f<3$. This implies that the orbifold projection of the parent ${\cal N}=2$ KK theory alone cannot produce a low energy spectrum with (the net number of) three chiral generations. We, therefore, would have to assume that there are some additional (``twisted'') sectors in the theory which give rise to the light modes corresponding to the rest of the $3+n^-_f-n^+_f$ chiral
families (and, possibly, to some additional ``vector-like'' families). Note, however, that in this setup the light modes from these new sectors do not have massive KK counterparts. (That is, the KK modes with the quark and lepton gauge quantum numbers arise only in the original sector
of the theory corresponding to the projected ${\cal N}=2$ KK theory.) We will discuss possible
sources of these additional sectors in section IV. Here, however, we will simply assume their existence, and study the gauge coupling unification in this model.

\section{Unification}

{}In this section we discuss the gauge coupling unification in the extension of the MSSM described in the previous section. To begin with, let us first consider the familiar case of 
the MSSM unification. Let $\alpha_a$, $a=1,2,3$, be the $U(1)_Y$, $SU(2)_w$ and $SU(3)_c$
gauge couplings, respectively. More precisely, we will use the standard normalization $\alpha_1={5\over 3}\alpha_Y$. The one-loop evolution of the gauge couplings is given by
\begin{equation}\label{running}
 \alpha^{-1}_a (\mu^\prime) = \alpha^{-1}_a (\mu) -{b_a\over 2\pi}\log {\mu^\prime\over\mu}~,
\end{equation}
where $b_a$ are the one-loop $\beta$-function coefficients in the MSSM:
\begin{equation}
 b_1={33\over 5}~,~~~b_2=1~,~~~b_3=-3~.
\end{equation}
If we take the LEP data for the gauge couplings at, say, $\mu=M_Z$ (where $M_Z$ is the $Z$-boson mass), then the extrapolation of $\alpha_a$ (in the assumption of the MSSM matter content above the supersymmetric thresholds $\sim 1~{\mbox{TeV}}$) leads to their unification
(within the experimental uncertainty) at $M_{\small{GUT}}\approx2\times 10^{16}~{\mbox{GeV}}$: $\alpha_1 (M_{\small{GUT}})\approx \alpha_2 (M_{\small{GUT}})
\approx \alpha_3 (M_{\small{GUT}})\approx\alpha_{\small{GUT}}\approx 1/24$. 

{}Now consider adding the new superfields $F_\pm$ to the light spectrum of the MSSM. Since
$F_\pm$ are neutral under $SU(3)_c\otimes SU(2)_w$, the corresponding $\beta$-function coefficients are unchanged. The $\beta$-function coefficient for $U(1)_Y$, however, is affected.
Let ${\widehat b}_a$ be the new $\beta$-function coefficients:
\begin{equation}\label{MSSM}
 {\widehat b}_1={39\over 5}~,~~~{\widehat b}_2=1~,~~~{\widehat b}_3=-3~.
\end{equation}
Since the new fields $F_\pm$ are ``vector-like'', {\em a priori} they can acquire non-zero mass
even if ${\cal N}=1$ supersymmetry is unbroken. We will refer to the mass scale of the superfields $F_\pm$ as $M_F$. Then in the one-loop logarithmic running of the gauge couplings    
we must use the $\beta$-function coefficients $b_a$ at energies below $M_F$, and ${\widehat b}_a$ at energies above $M_F$. Note that with this new matter content the unification prediction is ruined ({\em i.e.}, the three gauge couplings no longer meet at a single scale) unless $M_F$ is around $M_{\small{GUT}}$. 

{}However, the light spectrum\footnote{By ``light spectrum'' we mean the MSSM fields (whose masses are  ${\ \lower-1.2pt\vbox{\hbox{\rlap{$<$}\lower5pt\vbox{\hbox{$\sim$}}}}\ }M_{ew}$, where $M_{ew}$ is the electroweak scale), plus the new fields $F_\pm$ (whose mass scale $M_F$ {\em a priori} can be $M_{ew} {\ \lower-1.2pt\vbox{\hbox{\rlap{$<$}\lower5pt\vbox{\hbox{$\sim$}}}}\ } M_F
{\ \lower-1.2pt\vbox{\hbox{\rlap{$<$}\lower5pt\vbox{\hbox{$\sim$}}}}\ }
1/R$ - see below).} is accompanied by the infinite tower of the KK modes discussed in the previous section. Thus, above the KK threshold $1/R$ we must take into account the contributions of the KK modes to the gauge coupling running. Here we can proceed in two equivalent ways. We can ``integrate in'' the KK modes as we go to higher and higher energies.
The gauge coupling running is then no longer logarithmic but power-like above the KK threshold.
This way we can compute the modification to (\ref{running}) due to the KK modes as a function of the energy scale. Then we can see whether the running gauge couplings defined this way unify at some energy scale. Alternatively, we can assume that at some scale $M_s$ (which we will identify with the string scale in the brane world picture) the gauge couplings are the same, and compute the one-loop corrections to the gauge couplings at some low energy scale $\mu\ll1/R<M_s$. Then we can see whether the one-loop corrected gauge couplings at the energy scale $\mu$ agree with the experimental data.

{}Here we will follow the second approach. More precisely, in the string language we can assume that the tree-level gauge couplings are the same for all three gauge subgroups. The energy scale dependent one-loop corrections to the gauge couplings then come from the corresponding infra-red (IR) divergences in string theory \cite{kap}
(and the energy dependence arises via the IR cut-off), which give precisely the field theoretic logarithmic evolution of gauge couplings. There are also energy scale {\em independent} corrections from various one-loop {\em thresholds}. Thus, if we are interested in gauge couplings at energy scales $\mu\ll1/R$, these threshold corrections come from 
the heavy KK modes, as well as string states such as string oscillator modes. At one loop
we have 
\begin{equation}\label{running1}
 \alpha^{-1}_a (\mu) = \alpha^{-1} + {b_a\over 2\pi}\log \left({M_F\over\mu}\right)+
 {{\widehat b}_a\over 2\pi}\log\left({M_s\over M_F}\right)
 +\Delta_a~,
\end{equation}
where $\alpha=(2\pi)^{p-3} g_s/2M_s^{p-3} V_{p-3}$ is the ``unified'' gauge coupling at the 
string scale $M_s$, $g_s$ is the (ten dimensional) string coupling, and $V_{p-3}$ is the volume 
of the compact dimensions inside of the D$p$-brane world-volume. (Here we are using the conventions of \cite{pol}.) The threshold corrections $\Delta_a$ can be computed in string theory using the standard techniques \cite{kap}\footnote{The discussion in \cite{kap} applies to the perturbative heterotic superstring case. For a discussion of one-loop thresholds in the type I superstring context, see, {\em e.g.}, \cite{bac}.}. Here we are only going to be interested in the
thresholds due to the massive KK modes as they give the leading contributions in the regime
where $(RM_s)^{p-3}\gg 1$. In fact, using the results of \cite{TV}, we have (see appendix A for details):
\begin{equation}\label{thres}
 \Delta_a = {{\widetilde b}_a\over 2\pi} \eta_p (RM_s)^{p-3} 
 -{{\widetilde b}_a\over 4\pi} \log\left(R M_s\right) +{\cal O}(1)~,
\end{equation} 
where ${\cal O}(1)$ terms include the corrections due to string oscillator modes as well as other contributions. In (\ref{thres}) the $\beta$-function coefficients ${\widetilde b}_a$ are those of the
underlying ${\cal N}=2$ theory. In our examples this is the gauge theory given by the fields ${\widetilde \Phi}$. The $\beta$-function coefficients for the matter content discussed in section II are given by: 
\begin{equation}\label{N=2}
 {\widetilde b}_1={18\over 5}+4n_f~,~~~{\widetilde b}_2=-2+4n_f~,~~~{\widetilde b}_3=-6+4n_f~.
\end{equation}
Here $n_f$ is the number of ``flavors'' at the massive KK levels (see section II).

{}The appearance of the ${\cal N}=2$ $\beta$-function coefficients in the 
KK thresholds is not surprising as the massive KK modes have ${\cal N}=2$ supersymmetry. Also, the fact that in the $(RM_s)^{p-3}\gg 1$ regime the KK thresholds go as $\sim (RM_s)^{p-3}$
can be understood by noting that this is simply the number of the massive KK modes lying between the lowest KK threshold $1/R$ and $M_s$ (and each KK level contribution to $\Delta_a$ is of order 1). Finally, the coefficient $\eta_p$ is a (strictly speaking subtraction scheme dependent) numerical factor of order 1 (which we define more precisely in 
appendix A).  

{}The expression (\ref{running1}) is valid at energy scales below the KK threshold $1/R$. 
Suppose that $M_F$ is not much lower than $M_s$. Here we can ask whether the low energy couplings given by (\ref{running1}) agree with the experimental data. Using the fact that the gauge couplings unify at the scale $M_{\small{GUT}}$ in the MSSM context, it is not difficult to see that the condition that guarantees correct  values for $\alpha_a (\mu)$ in the present setup is the following:
\begin{equation}\label{nu}
 {\mbox{$\nu_{ab}$ is independent of $a,b$, where}}~
 \nu_{ab}\equiv {{\widetilde b}_a-{\widetilde b}_b
 \over b_a-b_b}~{\mbox{for $a\not=b$.}}
\end{equation} 
Recall that
$b_a$ are the MSSM $\beta$-function coefficients, whereas ${\widetilde b}_a$ are those
of the ${\cal N}=2$ gauge theory with ${\widetilde \Phi}$ superfields. Note that the above condition is satisfied in our model. In fact, we have
\begin{equation}\label{nu1}
 \nu_{ab}=1~~~\forall~a\not=b~.
\end{equation}
This, in particular, implies that
\begin{eqnarray}\label{uniA}
 &&\eta_p (RM_s)^{p-3}-{1\over 2} \log\left(R M_s\right) 
 \approx  \log \left({M_{\small{GUT}}\over M_s}\right)~,\\
 \label{uniA1}
 &&\alpha^{-1}\approx\alpha^{-1}_{\small{GUT}}-{\kappa\over 2\pi} \log\left( {M_{\small{GUT}}
 \over M_s}\right)~,
\end{eqnarray}
where 
\begin{equation}\label{kappa}
 \kappa\equiv{\widetilde b_a}-b_a=-3 +4n_f
\end{equation}
is independent of $a$. Here we would like to emphasize that if 
the mass scale $M_F$ of the $F_\pm$ superfields
is much lower than $M_s$, unification is no longer possible due to large contributions of the second logarithm in (\ref{running1}) as the ratios $({\widehat b}_a-{\widehat b}_b)/(b_a-b_b)$ do depend on $a,b$. Also, here we must assume that $M_F{\ \lower-1.2pt\vbox{\hbox{\rlap{$<$}\lower5pt\vbox{\hbox{$\sim$}}}}\ }1/R$. This is due to the following. A non-vanishing mass term for the light $F_\pm$ states generically implies that the entire tower of their massive KK counterparts is also shifted. 
This would affect the computation of the KK thresholds substantially if $M_F{\ \lower-1.2pt\vbox{\hbox{\rlap{$>$}\lower5pt\vbox{\hbox{$\sim$}}}}\ } 1/R$. To maintain control over the KK thresholds we, therefore, should require that $M_F$ is somewhat lower than $1/R$.

\subsection{Lower Bound on $M_s$}

{}In this subsection we discuss a lower bound on the unification scale $M_s$ arising in the model of section II. To see how this bound comes about, let us go back to the unification relations (\ref{uniA}),(\ref{uniA1}). In particular, let us examine the expression (\ref{uniA1})
for the ``unified'' gauge coupling $\alpha$, which, more precisely, is the tree-level gauge coupling.
It is clear that we must have $\alpha^{-1}>0$, which ultimately implies the following {\em lower} bound on $M_s$ (provided that $\kappa>0$):
\begin{equation}\label{bound}
 M_s {\ \lower-1.2pt\vbox{\hbox{\rlap{$>$}\lower5pt\vbox{\hbox{$\sim$}}}}\ }
  M_{\small{GUT}} \exp\left(-{2\pi\over\kappa\alpha_{\small{GUT}}}\right)~.
\end{equation}  
Thus, for $n_f=3$ we have $\kappa=9$, and the lower bound reads
(where we take $\alpha_{\small{GUT}}\approx1/24$, and $M_{\small{GUT}}\approx2\times 10^{16}~{\mbox{GeV}}$): $M_s{\ \lower-1.2pt\vbox{\hbox{\rlap{$>$}\lower5pt\vbox{\hbox{$\sim$}}}}\ }
10^9~{\mbox{GeV}}$. This implies that the mass scale $M_F$ of the new states $F_\pm$ cannot be much lower than $10^7-10^8~{\mbox{GeV}}$ (which is dictated by the fact that otherwise the unification prediction would essentially be ruined). {\em A priori} there is nothing inconsistent with having such large $M_F$, and such a scenario could be viable. In fact, since the states $F_\pm$ are ``vector-like'', generically there are two natural scales for $M_F$: either $M_F\sim M_X$, or $M_F\sim M_H$, where $M_H$ is the mass scale for the electroweak Higgs doublets. 
Here $M_X$ is some high scale which typically is of order $M_s$, but sometimes can be somewhat (albeit generically not too much) lower than $M_s$. For instance, $M_X$ can be related to an anomalous $U(1)$ scale (which can be an order of magnitude or so lower than $M_s$). Often it is actually more problematic to explain lower mass scales for such ``vector-like'' matter than the higher ones. In the case of the electroweak Higgs doublets this is known as the $\mu$-problem which can be posed as the question of why is the mass parameter $\mu$ in the $\mu$-term $\mu H_+ H_-$ of order of the electroweak scale $M_{ew}$ rather than, say, $M_s$ (provided that $M_s\gg M_{ew}$). A similar question could be asked about the ``$\mu^\prime$-term'' $\mu^\prime F_+ F_-$ provided that $M_F\sim M_H$.   

{}Nonetheless, since we expect the electroweak Higgs doublets not to be much heavier than $M_{ew}$, it is not unnatural to imagine a scenario where $M_F$ would also be of the same order of magnitude. On the other hand, for this to be compatible with the unification constraints
in the present context, we must assume that $M_s$ is not too much higher than TeV (since to preserve the unification prediction $M_s$ should not be too much higher than $M_F$). We can then wonder whether the unification scale can be lowered to such values.  From (\ref{bound}) it is clear that the only way to do so is to lower the value of $\kappa$. This is achieved by considering lower values of $n_f$.    

{}Note that for $n_f=0$ the bound (\ref{bound}) is absent as in this case $\kappa<0$, and the ``unified'' gauge coupling $\alpha$ is always smaller than $\alpha_{\small{GUT}}$. On the other hand, for $n_f\geq 1$ the bound (\ref{bound}) is non-trivial as $\kappa>0$ in these cases.
For $n_f=1$, however, the lower bound on $M_s$ comes out to be way below $M_{ew}$, so 
phenomenologically it is immaterial. For $n_f=2$  we have a non-trivial lower bound: $M_s{\ \lower-1.2pt\vbox{\hbox{\rlap{$>$}\lower5pt\vbox{\hbox{$\sim$}}}}\ } 1.6~{\mbox{TeV}}$. (Here we are neglecting contributions from possible additional ``vector-like'' generations in the light spectrum which would further raise the lower bound on $M_s$.)    

{}Thus, we see that in such a scenario $M_s$ could {\em a priori} be lowered to a few TeV. Then
the mass scale $M_F$ of the new light states $F_\pm$ could be as low as a few hundred GeV
(that is, around the mass scale of the electroweak Higgs doublets $H_\pm$). In such cases we have an experimental prediction - the new states $F_\pm$ beyond the MSSM spectrum - which
could be tested in the present or near future collider experiments.

\subsection{Higher Loop Corrections}

{}So far we have discussed one-loop renormalization of the gauge couplings. In this subsection we would like to address the issue of how reliable is the one-loop approximation. Naively, it might seem that as long as the ``unified'' gauge coupling $\alpha$ is small, the higher loop corrections are negligible. This is, however, not the case. The true expansion parameter
is {\em not} $\alpha/2\pi$ (which would be the case if the effective field theory description in terms of just the light modes was adequate all the way up to the string scale $M_s$). Rather, the correct expansion parameter is related to the string coupling $g_s$. This is due to the fact that although each KK mode (including the light modes) couples with the strength of order $\alpha$, there are many (namely, $(RM_s)^{p-3}$) KK modes between the KK threshold $1/R$ and the string scale $M_s$. All of these modes contribute into the gauge coupling renormalization at higher loops, and the true loop expansion parameter is enhanced\footnote{This is analogous to
considering the effective 't Hooft coupling in large $N$ gauge theories \cite{thooft}.} by a factor of $(RM_s)^{p-3}$ compared with $\alpha/2\pi$. 

{}Let us be more precise here. The loop expansion parameter is given by $\lambda_s N$ in the open string sector, and by $\lambda_s^2$ in the closed string sector. Here $\lambda_s\equiv
g_s/4\pi$ (as we have already mentioned, we are using the conventions of \cite{pol}). The factor of $N$ accompanying $\lambda_s$ in the loop expansion parameter for the open string sector is the number of D$p$-branes\footnote{Strictly speaking this factor at a given loop order is related to the corresponding $\beta$-function coefficient, but the rough estimate given by the number $N$ of D$p$-branes is going to be sufficient for our purposes here.}. Recall that $\alpha=(2\pi)^{p-3} g_s/2M_s^{p-3} V_{p-3}$, where $V_{p-3}=(2\pi R)^{p-3}/2$. This implies that
\begin{equation}
 \lambda_s ={\alpha\over 4\pi} (R M_s)^{p-3}~.
\end{equation}  
This expression makes it apparent that $\lambda_s$ need not be small (even if $\alpha$ is) for the factor $(RM_s)^{p-3}\gg 1$ if $M_s\ll M_{\small{GUT}}$. 

{}For illustrative purposes let us estimate $\lambda_s$ in some specific examples. 
Suppose the unification scale is $M_s\simeq
10~{\mbox{TeV}}$. Then we can estimate the corresponding compactification radius from
(\ref{uniA}). Here we need to specify the subtraction scheme to obtain $\eta_p$. For the sake of concreteness we will use the value of $\eta_p$ in the subtraction scheme used in \cite{dien}
(see footnote \ref{foot} in appendix A). For $p=5$ (that is, in the case of two extra compact dimensions inside of D5-branes) we then have $\eta_5=\pi/4$, and $RM_s\simeq 6.1$. Next, we can obtain the value
of $\alpha$ from (\ref{uniA1}) using (\ref{kappa}) for $\kappa$. Thus, for $n_f=1$ we have
$\alpha\simeq 1/19.5$, and $\lambda_s\simeq 0.15$. Similarly, for $n_f=2$ we have  
$\alpha\simeq 1/1.46$, and $\lambda_s\simeq 2.0$. Thus, we see that in the $n_f=2$ case the
loop expansion parameter is bigger than one. In fact, in the open string sector this is further enhanced by the number $N$ of D5-branes (which must be at least 4 to accommodate the Standard Model gauge group, but generically is expected to be even larger). Thus, in the $n_f=2$ case the perturbation theory breaks down for such low values of $M_s$. This, in particular, implies that we have no control over the higher loop corrections which could ruin the unification prediction. Note that $\lambda_s$ can be lowered by raising the unification scale $M_s$. Thus, to have, say, $\lambda_s{\ \lower-1.2pt\vbox{\hbox{\rlap{$<$}\lower5pt\vbox{\hbox{$\sim$}}}}\ } 0.15$, we must require
$M_s{\ \lower-1.2pt\vbox{\hbox{\rlap{$>$}\lower5pt\vbox{\hbox{$\sim$}}}}\ } 170~{\mbox{TeV}}$. 

{}Thus, we see that $n_f=0,1$ cases are preferred if we want to have relatively low values
of $M_s$ and, at the same time, maintain control over the higher loop corrections. In fact, even in these cases it is far from being obvious that the higher loop corrections do not ruin the unification prediction. Thus, let us go back to the case $n_f=1$ with $M_s\simeq 10~{\mbox{TeV}}$. In this case we have $\lambda_s\simeq 0.15$, but the loop expansion parameter in the open string (that is, gauge) sector of the theory is still expected to be (at least) of order 1 (which is due to the enhancement by a factor of $N$). However, as we will argue in a moment, the higher loop corrections are expected to be of order 1 (which are
subleading compared with the KK threshold corrections $\sim (RM_s)^{p-3}\gg1$)
provided that the loop expansion parameter is of order 1. As we will see, this statement is a non-trivial consequence of supersymmetry, and we would not have had any reason to expect it to hold in non-supersymmetric theories.

{}Thus, let us discuss the expected sizes of the higher loop corrections when the loop expansion parameter $\lambda_s N\sim 1$. To simplify the discussion, let us consider the case of a single gauge group $G$. (The generalization to a product gauge group is straightforward.). We will assume that the massless modes (with quantum numbers $\Phi$) are ${\cal N}=1$ supersymmetric, and the corresponding heavy KK modes (with quantum numbers ${\widetilde \Phi}=\Phi\oplus\Phi^\prime$) are ${\cal N}=2$ supersymmetric in complete parallel with our discussion in section II. Also, we will assume that the KK modes correspond to a single compact direction. (That is, here we are considering the case of D4-branes with one of the space-like directions inside of their world-volume compactified.)  
Let $\alpha$ be the tree-level gauge coupling. Furthermore, let $\alpha(\mu)$ be the renormalized gauge coupling at the scale $\mu$. 

{}On general grounds we expect the renormalized gauge coupling at scales $\mu\ll1/R$ to be given by 
\begin{equation}
 \alpha^{-1}(\mu)=\alpha^{-1} +f(\mu)+\Delta~,
\end{equation}
where $\Delta$ is the ($\mu$-independent) contribution due to the heavy KK thresholds, whereas $f(\mu)$ corresponds to the gauge coupling running. Here we are interested in estimating sizes of both $f(\mu)$ and $\Delta$. The $L$-loop order contribution $\alpha^{-1}_L(\mu)$ to $\alpha^{-1}(\mu)$ can be schematically written as
\begin{equation}\label{loop}
 \alpha^{-1}_L(\mu)=\alpha^{-1} \sum_{m=0}^{L} c_{m,L} (RM_s)^{m}\left(N\alpha\over   
 4\pi\right)^L~,
\end{equation}
where $m$ counts the number of the loop propagators corresponding to the heavy KK modes
with {\em independent} KK momentum summations. (Note that the KK momentum must be conserved inside of the loop diagram as the two external gauge boson lines carry zero KK momentum.) The enhancement factor $(RM_s)^{m}$ arises due to $RM_s$ heavy KK states propagating in each of these $m$ internal lines. The coefficients $c_{m,L}$ generically depends on the energy scale $\mu$ via an appropriate IR cut-off. Note that for $RM_s\gg1$ the coefficients $c_{m,L}{\ \lower-1.2pt\vbox{\hbox{\rlap{$<$}\lower5pt\vbox{\hbox{$\sim$}}}}\ } 1$ for energy scales $\mu$ not too much smaller than $1/R$. (Thus, here and in the following we will treat logarithms of the type $\log(RM_s)$ as being of order 1.)

{}For $\lambda_sN\sim 1$, naively one might expect a large (that is, of order $\alpha^{-1}$) contribution to $\alpha^{-1}_L(\mu)$ coming from the term $m=L$ in (\ref{loop}). (Note that this term is $\mu$-independent. More precisely, it is finite in the limit $\mu\rightarrow 0$.) However, as we will show in a moment, for $L\geq 2$ this contribution is actually suppressed by an additional power of $RM_s$, and, therefore, it is at most of order $N$. This is much smaller than the corresponding one-loop contribution (due to the one-loop KK threshold corrections)
which is of order $N(RM_s)$ as can be seen from (\ref{thres}). Thus, we have    
\begin{eqnarray}
 &&(N\alpha_1(\mu))^{-1}\sim RM_s~,\\
 &&(N\alpha_{L>1}(\mu))^{-1}\sim 1~,
\end{eqnarray}
where the estimates here should be understood symbolically (that is, we are suppressing the $\mu$ dependence in the corresponding contributions). This implies that higher loop contributions to the gauge coupling renormalization are subleading compared with the one-loop contribution, albeit they are of order 1 so we have to worry about the perturbative ``convergence'' issues (which we will discuss in a moment). However, first let us show that the higher loop contributions are indeed of order 1 instead of $RM_s$.

{}The key observation here is the following. Consider the gauge coupling renormalization in the parent ${\cal N}=2$ gauge theory with the gauge group $G$ (with the same tree-level gauge coupling $\alpha$ as in the ${\cal N}=1$ theory). In this theory the massless as well as heavy KK modes carry ${\widetilde \Phi}$ quantum numbers. The number of the heavy KK modes, however, is twice that in the ${\cal N}=1$ theory (which is due to the ${\bf Z}_2$ orbifold projection in the latter). For the ${\cal N}=2$ theory we have:
\begin{equation}\label{loop2}
 \alpha^{-1}_L(\mu)=\alpha^{-1} \sum_{m=0}^{L} {\widetilde c}_{m,L}    
 (RM_s)^{m}\left(N\alpha\over   
 2\pi\right)^L~,
\end{equation}
where we have $(N\alpha/2\pi)^L$ instead of $(N\alpha/4\pi)^L$ (as in (\ref{loop})) as the loop expansion parameter in the ${\cal N}=2$ theory is $N{\widetilde \lambda}_s=N\alpha(RM_s)/2\pi=2N\lambda_s$. (Recall that in the ${\cal N}=2$ theory the volume of the compactified direction inside of the D4-branes is twice that in the ${\cal N}=1$ theory.) In fact, with this normalization we have ${\widetilde c}_{L,L}=c_{L,L}$. Now, the gauge coupling is not renormalized beyond one loop in the ${\cal N}=2$ theory: $\alpha^{-1}_{L>1}(\mu)\equiv 0$. This implies that due to ${\cal N}=2$ supersymmetry there is a cancellation in the coefficient ${\widetilde c}_{L,L}$ (for $L>1$) such that it is at most as large as $1/RM_s$ (or else the required cancellation between the $m=L$ term and the terms with $m<L$ would not be possible). Thus, we have shown that $c_{L,L}={\cal O}(1/RM_s)$ for $L>1$, which in turn implies that in the ${\cal N}=1$ theory we indeed have $(N\alpha_{L>1}(\mu))^{-1}\sim 1$.

{}Having established that in the ${\cal N}=1$ theory the leading correction comes from the one-loop KK thresholds, we must now worry about the fact that all the higher loop corrections are actually of order 1 (for $\lambda_s N\sim 1$). Strictly speaking we cannot even trust the perturbation theory. However, it appears that supersymmetry once again offers a way out this\footnote{I would like to thank Nima Arkani-Hamed and Gia Dvali for discussions on this point.}. We can gain control over the higher loop corrections by considering the {\em holomorphic} gauge coupling $\alpha(\mu)$ which in the ${\cal N}=1$ theory runs only at one loop \cite{SV}. In fact, we have the following relation \cite{SV}:   
\begin{equation}
 \alpha^{-1}(\mu)=\alpha^{-1}+\alpha^{-1}_1 (\mu) + {1\over 2\pi}\sum_i T_i \log
 Z_i (\mu)~,
\end{equation}
where $\alpha^{-1}_1 (\mu)$ is the one-loop contribution to $\alpha^{-1}(\mu)$, 
$T_i$ are the $\beta$-function coefficients ($T_i\delta^{ab}={\mbox{Tr}}_i (T^a T^b)$) for {\em massless} chiral supermultiplets (or, more precisely, light chiral supermultiplets with masses $\ll1/R$), and $Z_i (\mu)$ are the corresponding wave function renormalization coefficients. What we would like to know is the size of the $\log Z_i$ contributions, in particular, if it can be comparable with the one-loop heavy KK threshold contribution (which is of order $N(RM_s)$). This seems to be unlikely as this would imply an exponentially large wave function renormalization in a theory with an order 1 ``effective coupling'' $\lambda_s N$, so we can expect that $\log Z_i$ are of order 1. In other words, intuitively the two {\em a priori} natural values for $\log Z_i$ (which are not $\ll 1$) are $\sim RM_s$ or $\sim 1$. (Recall that we treat quantities like $\log (RM_s)$ as $\sim 1$.) The first possibility is hard to imagine as $RM_s$ appears in $\log Z_i$ only via the combination $\lambda_s N\propto N\alpha (RM_s)$.   

{}To make the above argument more precise, let us consider the perturbative expansion for $\log Z_i$. On general grounds we expect it to have the following form:
\begin{equation}\label{Z}
 \log Z_i (\mu) =\sum_{L=1}^\infty \sum_{m=0}^L d_{m,L} (RM_s)^m\left({N\alpha\over 4\pi}\right)^L~,
\end{equation}
where the coefficients $d_{m,L}$ have properties similar to those of $c_{m,L}$. The above expression implies at a given loop order the leading contribution comes from the $m=L$ term, and it is of order\footnote{Here we should point out the following. Consider the case where the number of ``flavors'' $n_f$ at the heavy KK levels is 3, and the light chiral generations arise solely via the orbifold reduction of the parent ${\cal N}=2$ theory (that is, there are no additional sectors giving rise to chiral (or ``vector-like'') generations.) In this case the couplings of all of the light chiral superfields to the heavy KK modes are dictated by the parent ${\cal N}=2$ supersymmetry. This, in particular, implies that the coefficients $d_{L,L}$ are not of order 1 in this case, but of order $1/RM_s$. Thus, in this case we expect $\log Z_i\sim 1/RM_s$, which would imply that we have control over the Yukawa coupling renormalization (see below). (Note, however, that the perturbative expansion parameter for $\log Z_i$ is still of order one, so we still need the ``convergence'' argument given below for the general case.) However, as we discuss in the next section, in this case the Yukawa couplings are not realistic, and there is no top-like generation. On the other hand, if we have additional chiral families coming from the sectors with no heavy KK counterparts, $d_{L,L}$ is no longer suppressed as these states generically have ${\cal N}=1$ (and not ${\cal N}=2$) couplings with the heavy KK modes in the orbifold reduction of the parent ${\cal N}=2$ theory.} $(\lambda_s N)\sim 1$. Thus, we see that the leading contribution to the holomorphic gauge coupling indeed comes from the one-loop heavy KK thresholds. However, the issue here is whether the series (\ref{Z}) is convergent. In certain cases this series can be resummed \cite{Jones,Ross}, and we expect to have a finite convergence radius\footnote{This convergence radius depends on the details of the model such as the explicit form of the superpotential.} $\xi_c\sim 1$, so that for $\lambda_s N<\xi_c$ the series is convergent. Moreover, for $\lambda_s N$ within the convergence radius and such that $\xi_c - \lambda_s N\sim 1$ the value of the resummed series is expected to be of order 1 \cite{Jones,Ross} in accord with the above intuitive arguments. In fact, it is not unreasonable to believe that non-perturbatively this might also be the case as long as $\lambda_s N\sim 1$. 

{}The above considerations suggest that the leading contribution to the renormalization of the holomorphic gauge coupling is indeed coming from the one-loop heavy KK thresholds, and the unification prediction is likely to persist even for $\lambda_s N\sim 1$. That is, we might be able to neglect (with the uncertainty of order $1/(RM_s)^{p-3}$ in the case of D$p$-branes) the higher loop contributions as long as $\alpha$ is small and $\lambda_s N\sim 1$ (or, at least, if $\lambda_s N<\xi_c$).

{}Here we would like to point out that even though the unification prediction for the holomorphic gauge couplings may be safe even for $\lambda_s N\sim 1$, we have no way of arguing the same for Yukawa couplings. Thus, $Z_i$ are related to the Yukawa coupling renormalization, and, apart from the fact that they are likely to be of order 1, {\em a priori} we have no reason to believe that, say, any one-loop unification prediction for Yukawa couplings would persist at higher orders as well. 

{}Finally, the entire discussion of this subsection heavily relies on supersymmetry. Suppose now we have $M_s$ of order a few TeV, and $1/R$ of order several hundred GeV. Since the supersymmetry breaking scale should be right around these scales, we would lose control over the KK tower, the corresponding thresholds, and higher loop corrections which could now go haywire. This suggests that perhaps the string scale should be pushed to $10-100~{\mbox{TeV}}$ to have a safe unification prediction in this type of models. In fact, this might also be desirable from other points of view: clearly, it should be easier to accommodate various experimental bounds with a slightly higher string/KK threshold scale. Moreover, a little room between the string scale and supersymmetry/electroweak breaking scale might be useful for (perhaps, dynamically) generating various hierarchies (such as fermion flavor hierarchy \cite{flavor}) in the Standard Model.     

\section{Brane World Embedding}

{}In the previous sections we have proposed an extension of the MSSM where gauge coupling unification can occur at scales (much) lower than $M_{\small{GUT}}$. One of the key ingredients
of our model is the presence of extra dimensions which make it possible to lower the unification scale $M_s$ along the lines of \cite{dien}. On the other hand, the fact that (at one loop) the gauge couplings unify in this model just as precisely as in the MSSM is due to its matter content. 
In particular, the massive KK excitations of the new fields $F_\pm$ are
crucial for obtaining the ${\cal N}=2$ $\beta$-function coefficients
${\widetilde b}_a$ with the desired properties\footnote{The fact that the
unification constraint (\ref{nu}) is satisfied in our model is
non-trivial. Thus, in the models of \cite{dien} 
(which do not appear to be embeddable in the brane world framework) 
this constraint is satisfied only approximately so that the actual predictions for the low energy gauge couplings and, say, $\sin^2(\theta_W)$ are expected to be a bit off.}. 

{}In this section we would like to discuss certain issues concerning possible embeddings of these models in string theory. More precisely, our discussion will be in the context of the brane
world scenario \cite{BW}, where the Standard Model fields are localized on some D$p$-branes
(with $p>3$ to allow for extra dimensions discussed above), whereas gravity propagates in a larger\footnote{Note that in string theory we expect both gauge and gravitational couplings to unify at the string scale $M_s$. In the context of, say, Type I string theory this appears to require $p<9$ \cite{BW}.} (10 or 11) dimensional bulk. Here we can ask whether it is possible to embed our model, or at least some of its features, in the brane world context.

{}The above question has two sides to it. First, we can ask whether we can {\em explicitly} construct a consistent string vacuum which reduces to our model at low energies. At present it is unknown how to do this, but this might be due to the lack of necessary technology as
many consistent string vacua have not yet been understood. Thus, it is still unknown how to embed, say, the MSSM (with no extra matter charged under $SU(3)_c\otimes SU(2)_w\otimes U(1)_Y$) in string theory. On the other hand, in our model we require some additional features
compared with the MSSM (or its four dimensional ${\cal N}=1$ supersymmetric extensions). In particular, to obtain the desired KK spectra we start with an ${\cal N}=2$ theory and then orbifold it. What we would like to discuss next is how such orbifold actions could be embedded in the brane world framework.

\subsection{Voisin-Borcea Orbifolds}

{}One of the simplest ways of obtaining models with some of the features discussed in the previous sections is via Type I (or Type I$^\prime$)\footnote{For recent developments in four dimensional Type I (Type I$^\prime$) compactifications/orientifolds, see, {\em e.g.}, \cite{typeI,KST,3gen}.} compactifications on Voisin-Borcea orbifolds \cite{Voisin}. Here we would like to review some facts about these Calabi-Yau three-folds. Let ${\cal W}_2$ be a K3 surface (which is not necessarily an 
orbifold) which admits an involution $J$ such that it reverses the sign of the holomorphic
two-form $dz_1\wedge dz_2$ on ${\cal W}_2$. Consider the following quotient:
\begin{equation}
 {\cal Y}_3= (T^2\otimes {\cal W}_2)/Y~,
\end{equation}
where $Y=\{1,S\}\approx {\bf Z}_2$, and $S$ acts as $Sz_0=-z_0$ on $T^2$ 
($z_0$ being a complex coordinate on $T^2$), and as $J$ on ${\cal W}_2$. This
quotient is a Calabi-Yau three-fold with $SU(3)$ holonomy which is elliptically 
fibered over the base ${\cal B}_2={\cal W}_2/B$, where 
$B=\{1,J\}\approx{\bf Z}_2$.   

{}Next, consider Type I compactification on a Voisin-Borcea orbifold ${\cal Y}_3$. Here one needs to specify the gauge bundle embedded in the D9-brane gauge group. For certain choices of the gauge bundle as well as the base ${\cal B}_2$, we can also have D5-branes. Here we are going to be interested in D5-branes wrapped on the fibre $T^2$. The low energy four dimensional gauge theory living in the world-volume of these D5-branes has ${\cal N}=1$ supersymmetry. We also have KK states corresponding to compactifying the original six dimensional D5-brane world-volume theory on $T^2$. More precisely, we can think about these KK modes as follows. First consider Type I compactified on K3. This six dimensional theory has ${\cal N}=1$ supersymmetry. Next, consider D5-branes whose transverse directions correspond to K3. That is, these D5-branes fill the six dimensional Minkowski space ${\bf R}^{5,1}$. Let us now further compactify two of the directions in ${\bf R}^{5,1}$ on a two-torus $T^2$. The corresponding four dimensional theory has ${\cal N}=2$ supersymmetry. Thus, the low energy effective theory of the D5-brane world-volume theory is an ${\cal N}=2$ gauge theory in four dimensions. The corresponding KK excitations also have ${\cal N}=2$ supersymmetry from the four dimensional viewpoint. Now let us orbifold this theory by the ${\bf Z}_2$ orbifold group $Y$ whose generator acts as a reflection $z_0\rightarrow -z_0$ on $T^2$, and as the involution $J$ on K3. The latter breaks the $SU(2)_R$ R-parity group of the ${\cal N}=2$ gauge theory to $U(1)_R$. Correlated together with the reflection of $T^2$, it produces an ${\cal N}=1$ supersymmetric gauge theory in four dimensions plus the KK excitations which still come in ${\cal N}=2$ supermultiplets. In fact, this is precisely the ${\bf Z}_2$ orbifold action described in section II.      

{}Here the following remark is in order. Perturbatively, when considering Type I compactifications on orbifolds, the orbifold group action on the Chan-Paton factors is typically fixed by the tadpole cancellation conditions. More precisely, suppose we view a given Type I compactification as a Type IIB orientifold. Then generically the action of the orientifold on the Chan-Paton factors is determined (or at least severely constrained) by the one-loop tadpole cancellation conditions. As an example let us consider the case where K3 in the above construction is a ${\bf Z}_2$ orbifold limit of $T^4$: ${\cal W}_2=T^4/{\bf Z}_2$. The corresponding Voisin-Borcea orbifold then has the following geometry: ${\cal Y}_3=(T^2\otimes T^4)/{\bf Z}_2\otimes {\bf Z}_2$. Let the generators of the two ${\bf Z}_2$'s be $g_1$ and $g_2$.
The action of these orbifold elements on the Chan-Paton factors is given by $N\times N$ matrices (where $N$ is the number of D-branes of the corresponding type). These matrices must form a (projective) representation of (the double cover of) the orbifold group ${\bf Z}_2\otimes {\bf Z}_2$. The perturbative tadpole cancellation conditions require that the Chan-Paton matrices $\gamma_{g_1}$ and $\gamma_{g_2}$ corresponding to the elements $g_1$ respectively $g_2$ are traceless: ${\mbox{Tr}}(\gamma_{g_{1,2}})=0$. This implies that both of the ${\bf Z}_2$ subgroups act non-trivially on the gauge quantum numbers of, say, the D5-branes in this example. In fact, typically this is the case in perturbative orientifolds. On the other hand, in our construction in section II we have assumed that the ${\bf Z}_2$ orbifold action on the gauge quantum numbers was trivial. This might appear as a shortcoming of our construction is section II, but here one should keep in mind that for {\em non-perturbative} orientifolds a non-trivial action of the orbifold group on the Chan-Paton factors may not be required. In fact, in the following we will argue that if there is an orientifold embedding for our model it should be non-perturbative.     
     
\subsection{Twisted Sectors}

{}Recall from section III that in order to lower the string scale $M_s$ below $10^9~{\mbox{GeV}}$ we had to assume $n_f<3$, where $n_f$ is the number of ``flavors'' at the massive KK levels.  As we discussed in section II, in such cases some of the three chiral families in the light spectrum must come from some additional sectors such that light states from these sectors have no KK counterparts. It is natural to ask whether such states can arise in string theory. The answer to this question is positive in the following sense. Since we are considering orbifold compactifications, the corresponding quotients will have a set of points fixed under the action of the orbifold group. For instance, in the case of the Voisin-Borcea orbifolds the ${\bf Z}_2$ twist $S$ (see the previous subsection) can act with fixed points. In particular, the fixed point set has real dimension two in this case. At each fixed point there is a collapsed two-sphere ${\bf P}^1$. If the world-sheet description is adequate D-branes wrapped on such ${\bf P}^1$'s do not give rise to non-perturbative states (which is due to the B-flux stuck inside of the ${\bf P}^1$'s \cite{aspin}).      
However, as was argued in \cite{KST}, for certain compactifications we must turn off the B-flux, and wrapped D-branes do give rise to non-perturbative ``twisted'' sector states. Note that in the case of the Voisin-Borcea orbifolds in the present context these states would live in the non-compact ${\bf R}^{3,1}$ part of the D5-brane world-volume, but they are localized at points on the fibre $T^2$ which the D5-branes wrap. This implies that such light twisted sector states do not have KK counterparts corresponding to the fibre $T^2$. Thus, to obtain the additional chiral sectors with the lepton and quark quantum numbers we can consider certain {\em non-perturbative} orientifolds with twisted sectors. As explained in detail in \cite{KST}, such sectors do not possess world-sheet description. One way to understand such states is to consider the map \cite{sen} of (the T-dual of) the corresponding Type I vacuum to F-theory \cite{vafa}. (Such a map for the Voisin-Borcea orbifolds was discussed in detail in \cite{KST}.) 

{}Thus, in non-perturbative orientifold constructions there might be twisted sector states with some of the properties discussed in section II. Here we would like to point out the following. Consider such a non-perturbative compactification on a Voisin-Borcea orbifold. Geometrically it is clear that there are going to be (a multiple of) 4 fixed points (of $T^2/{\bf Z}_2$). Thus, if there are any chiral families coming from such fixed points, they will always appear in multiples of 4. (In certain cases it might be possible to reduce this multiplicity to 2 by considering Type IIA orientifolds with D4-branes so that the relevant orbifold is $S^1/{\bf Z}_2$.) Here we are not particularly concerned with the multiplicity of such degenerate families. Rather, they will have identical Yukawa couplings with the electroweak Higgs doublets (which arise in the perturbative ``untwisted'' sectors). This is not very appealing for the following reason. It is not difficult to see that there are no renormalizable Yukawa couplings between the Higgs doublets and the ``untwisted'' sector generations. This simply follows from the fact that these states arise via the corresponding orbifold reduction of the parent ${\cal N}=2$ theory where such couplings are absent. This implies that to have a top-like generation we must have ``twisted'' sectors. However, the trouble here is that we will have more than one top-like generations coming from such twisted sectors as the number of the corresponding fixed points is always a multiple of two. (In fact, the corresponding bottom quarks as well as all the other quarks and leptons within these ``twin'' generations would also be degenerate.) This would make the entire scheme phenomenologically inviable.      

{}Here we can think of two ways of possibly circumventing the above difficulty. First, we can turn on non-trivial Wilson lines on the fibre $T^2$. These Wilson lines then can discriminate between the four fixed points of $T^2/{\bf Z}_2$ as the gauge quantum numbers of matter fields charged under $SU(3)_c\otimes SU(2)_w\otimes U(1)_Y$ can be different \cite{INQ}. More precisely, as we explain in a moment, the Wilson lines should act trivially on the $SU(3)_c\otimes SU(2)_w\otimes U(1)_Y$ gauge quantum numbers, that is, the corresponding string model should contain the $SU(3)_c\otimes SU(2)_w\otimes U(1)_Y$ gauge subgroup {\em  before} we turn on the Wilson lines. The Wilson lines can, however, act non-trivially on other gauge quantum numbers (corresponding to hidden and/or horizontal gauge symmetries), which could project out some of the original chiral generations and/or project in new ones with different quantum numbers under the other gauge subgroups. This way different fixed points can give rise to chiral generations with different couplings to the electroweak Higgs doublets (or, alternatively, we could imagine having only one twisted sector chiral generation with a large Yukawa coupling to the Higgs doublets). In fact, Wilson lines might even be necessary to obtain models with, say, (the net number of) three chiral families. Before we finish this discussion of possibly using Wilson lines, let us explain why the $SU(3)_c\otimes SU(2)_w\otimes U(1)_Y$ gauge subgroup should be present before turning on Wilson lines. An alternative would be to have a larger gauge group $G\supset SU(3)_c\otimes SU(2)_w\otimes U(1)_Y$ which is broken to $SU(3)_c\otimes SU(2)_w\otimes U(1)_Y$. It is not difficult to check that in this case the ${\cal N}=2$ $\beta$-function coefficients entering in the expression (\ref{thres}) for the heavy KK thresholds are no longer given by ${\widetilde b}_a$ defined in (\ref{N=2}), but rather by new $\beta$-function coefficients ${\widetilde b}_a^\prime$. The latter are computed as follows. Take the ${\cal N}=2$ gauge theory with the gauge group $G$ and the corresponding massless matter. Decompose each massless representation (including vector supermultiplets and hypermultiplets) according to the branchings under the breaking $G\supset SU(3)_c\otimes SU(2)_w\otimes U(1)_Y$. Compute the one-loop $\beta$-function coefficients for the $SU(3)_c$, $SU(2)_w$ and $U(1)_Y$ subgroups. This way one obtains ${\widetilde b}_3^\prime$, ${\widetilde b}_2^\prime$ and ${\widetilde b}_1^\prime$. Clearly, generically ${\widetilde b}_a^\prime\not={\widetilde b}_a$. (For instance, if $G$ is simple (such as $SU(5)$), then all three $\beta$-function coefficients ${\widetilde b}_a^\prime$ are identical.) The corresponding heavy KK thresholds then do not give rise to the correct prediction for the low energy gauge couplings.

{}The second possibility for avoiding more than one top-like generations is to consider ${\bf Z}_N$ orbifolds with $N\not=2$. Thus, in analogy with the Voisin-Borcea orbifolds, we can consider Calabi-Yau three-folds with $SU(3)$ holonomy given by the quotients 
\begin{equation}
 {\cal Y}_3=(T^2\otimes {\cal W}_2)/{\bf Z}_N~,
\end{equation}
where ${\cal W}_2$ is a K3 surface. Note that ${\bf Z}_N$ must act crystallographically on $T^2$, which implies that we can only have $N=2,3,4,6$. (The $N=2$ case corresponds to the Voisin-Borcea orbifolds.) In the ${\bf Z}_3$ case we have 3 fixed points of $T^2/{\bf Z}_3$. Thus, here we will still need Wilson lines to discriminate between the fixed points. In the ${\bf Z}_4$ case the number of points of $T^2$ fixed under ${\bf Z}_4$ is multiple of 2, so Wilson lines would be required here as well. However, in the case of ${\bf Z}_6$ we have only one point on $T^2$ which is fixed under the action of the generator $g$ of ${\bf Z}_6$. Thus, the ${\bf Z}_6$ twisted sectors (that is, the $g$ and $g^5$ twisted sectors) may give rise to a top-like generation. Other (that is, ${\bf Z}_2$ ({\em i.e.}, $g^3$) and ${\bf Z}_3$ ({\em i.e.}, $g^2$ and $g^4$)) twisted sectors in this ${\bf Z}_6$ orbifold may also give rise to light generations (this depends on the choice of the gauge bundle), but their couplings with the electroweak Higgs doublets need not be the same as that of the chiral generation coming from the ${\bf Z}_6$ twisted sectors.

{}Finally, we would like to make the following remark. The action of the ${\bf Z}_2$ (and, more generally, ${\bf Z}_N$) orbifold on the gauge quantum numbers is assumed to be trivial in the above discussions. (More precisely, this action is trivial on the quantum numbers under the $SU(3)_c\otimes SU(2)_w\otimes U(1)_Y$ gauge subgroup which must be present before orbifolding.) Here we can ask whether we could relax this requirement. The non-trivial issue that arises here is the same as in the discussion of the Wilson lines: if the orbifold group breaks some larger gauge symmetry $G$ down to $SU(3)_c\otimes SU(2)_w\otimes U(1)_Y$, then the $\beta$-function coefficients ${\widetilde b}_a^\prime$ appearing in the KK thresholds   
are generically different from ${\widetilde b}_a$. Can we find a non-trivial example such that $({\widetilde b}^\prime_a - {\widetilde b}^\prime_b)/({\widetilde b}_a - {\widetilde b}_b)$ is independent of $a,b$? (This would guarantee correct values of the one-loop renormalized low energy gauge couplings.) First note that the choices for the gauge group $G$ are rather limited: it has to be a subgroup of $E_6$, and it cannot be simple. (Moreover, the $SU(3)_c\otimes SU(2)_w$ cannot be a subgroup of a simple group which is a subgroup of $E_6$.) Thus, we can try $SU(4)_c\otimes SU(2)_w\otimes U(1)$, $SU(4)_c\otimes SU(2)_w\otimes SU(2)$, or
$SU(3)_c\otimes SU(3)_w\otimes SU(3)$ subgroups\footnote{Here we would like to point out that light states with the $F_\pm$ quantum numbers can arise in the ``trinification'' scenarios with the $SU(3)_c\otimes SU(3)_w\otimes SU(3)$ gauge group. They can come from the Higgs chiral superfields in $({\bf 1}, {\bf 3}, {\overline {\bf 3}})\oplus ({\bf 1}, {\overline {\bf 3}}, {\bf 3})$ 
of $SU(3)_c\otimes SU(3)_w\otimes SU(3)$ which are required to break the latter down to
$SU(3)_c\otimes SU(2)_w\otimes U(1)_Y$. For a toy brane world embedding of the $SU(3)_c\otimes SU(3)_w\otimes SU(3)$ gauge group with three chiral families, see \cite{KaSi}.}
of $E_6$. We have not been able to find (simple) matter contents for these gauge groups giving rise to the $\beta$-function coefficients ${\widetilde b}_a^\prime$ with the desired properties. This implies that the problem is rather restrictive, and the non-trivial solution we have found in this paper in the context of the Standard Model gauge group $SU(3)_c\otimes SU(2)_w\otimes U(1)_Y$ is not just ``one out of many''.

\section{Summary and Open Questions}

{}Let us summarize the discussions in the previous sections. We have considered an extension of the MSSM whose light spectrum contains new fields $F_\pm$ in addition to the MSSM fields, and there also are ${\cal N}=2$ supersymmetric Kaluza-Klein modes corresponding to the wrapping of D$p$-branes on $(S^1)^{p-3}$. The (one-loop) gauge coupling unification in our model occurs with just as good precision as in the MSSM. The unification scale depends on the details of the spectrum (more precisely, on the number $n_f$ of the ``flavors'' of quarks and leptons at heavy KK modes), as well as on the compactification radius $R$ (of the $S^1$'s). For instance, if $n_f=1$ we have $M_s\simeq 10~{\mbox{TeV}}$ provided that $(RM_s)^{p-3}\simeq 37$. In the case of D4-branes this would imply that the radius $R$ would have to be approximately 40 times inverse $M_s$. This is a relatively large number, and one might wonder whether it is ``natural''. However, for such low values of $M_s$ to have the correct value for the four dimensional Planck scale some of the compactified directions transverse to the D$p$-branes must be at least several orders of magnitude larger than $1/M_s$, so a factor of 40 does not seem so bad compared with the problem of the ``hierarchy'' of scales that the TeV string scenario faces. At any rate, in the case of D5-branes we have $R\simeq 6/M_s$, which might appear more appealing. 

{}Note that the unification at low scales such as $10-100~{\mbox{TeV}}$ via the heavy KK thresholds ultimately implies that the string coupling $g_s$ is not small, in fact, it is of order one
(more precisely, the open string loop expansion parameter $\lambda_s N\sim 1$). On the one hand, this might appear disappointing as the corresponding string vacuum is strongly coupled, and one can no longer enjoy the benefits of perturbation theory. On the other hand, in string theory dilaton stabilization appears to imply that the corresponding string vacuum must have string coupling of order 1 \cite{BaDi,BW}, and all the known mechanisms of dilaton stabilization
\cite{kras} also seem to support this conclusion. So perhaps the fact that the string coupling in the unification scenarios of this type is predicted to be of order 1 should be taken with moderate optimism.

{}Here we would like to emphasize that the particular model that we
considered in this paper is not ``generic'' in the sense that at present we
are not aware of any other models of this type where the gauge coupling
unification ``works'' as well as in the MSSM. At first this might appear as
a shortcoming, but there is a sense in which one can argue the
opposite. Thus, any attempt to lower the unification scale below
$M_{\small{GUT}}$ (which is the prediction of the MSSM within the
``desert'' assumption) will always raise the following question: Is the
unification within the MSSM a complete accident, or can it be explained by
the new scenario which has the pretense of replacing the old framework? It
would be nice if the latter were true. In our model this appears to be the
case in the following sense. Basically, our model tells us that the gauge
coupling unification in the MSSM is not an accident at all, but can be
explained by the lack of experimental data. Thus, suppose for a moment that
our model were (a part of) the correct description of nature above the
electroweak scale. Then a theorist who does not know that there exist
states $F_\pm$ as well as the entire tower of the KK modes with extended
(that is, ${\cal N}=2$) supersymmetry but does know (as an experimental
fact) about the existence of all the MSSM fields (except for the
electroweak Higgs doublets $H_\pm$ whose existence is assumed for the
standard radiative electroweak breaking mechanism to go through) will
sooner or later find that the gauge couplings in the MSSM context unify. 
(Here we are taking ${\cal N}=1$ supersymmetry for granted.)
The same theorist might be a bit surprised, however, when the new light (with masses of order of, say, a few hundred GeV) states $F_\pm$ are discovered but the heavy KK modes are still inaccessible to a direct experimental detection. Only after the latter are also found the ``true'' unification scheme would become clear.

{}The above discussion indicates that discovery of light $F_\pm$ states might have at least an indirect implication for our understanding of coupling unification. Our model predicts that if $F_\pm$ are light enough (meaning that they are accessible to the present or near future collider experiments) the string scale should also be ``around the corner''. On the other hand, there is always a possibility that they are rather heavy, in which case our model would predict higher values of $1/R$ and $M_s$ which would most likely not be observed in the near future directly in the high energy scattering experiments.

{}Next, we would like to discuss the open questions surrounding the issue of ``TeV scale'' coupling unification. The most important issue, we believe, might be the lack of an explicit string construction of a vacuum with the desired properties. As we mentioned before, this might be due to the lack of necessary technology. Thus, as we pointed out in section IV, if our model is realized as an orientifold (or a Type I/Type I$^\prime$ compactification), it should be non-perturbative (not only in the sense that the corresponding string coupling is of order one, but also that some of the light states should come from wrapped D-brane sectors). Perturbative orientifolds in four dimensions with ${\cal N}=1$ supersymmetry have been understood rather fully by now \cite{typeI,KST,3gen}. In fact, ``semi-realistic'' models with three chiral generations of quarks and leptons were obtained in this frameworks in \cite{3gen}. Non-perturbative orientifolds still need to be understood more completely, albeit some preliminary steps in these directions have been made in \cite{Alda,KST}, and the framework (which makes use of various dualities) for making further progress has been set up in \cite{KST}. In any case, better understanding of Type I/Type I$^\prime$, orientifold as well as F-theory compactifications to four dimensions is more than desirable.

{}Here we should point out that the heterotic M-theory framework \cite{HW} also might be interesting to consider in the present context. Thus, in \cite{ovrut} some progress has been made in understanding and formulating the systematic rules for constructing ${\cal N}=1$ supersymmetric heterotic M-theory vacua in four dimensions. This should facilitate model building in this context. It would, in particular, be interesting to see whether the (wrapped) M5-branes can accommodate the Standard Model. If so, one could attempt to lower the unification scale in this context as well. (Note that these vacua would also fall under the general category
of the brane world picture.)

{}Finally, we should point out that there might be other ways of solving the unification problem in the TeV string context. One such scenario was suggested in \cite{ST}. Thus, imagine that the Standard Model fields live in the world-volume of not one set of parallel D-branes but rather in the intersection of different D-branes. In fact, the ``bifundamental'' structure of the Standard Model (or the MSSM) matter content bodes well with this picture provided that, say, $SU(3)_c$ and $SU(2)_w$ (and even $U(1)_Y$) live in the world-volumes of different branes. The gauge couplings of these gauge subgroups would then be determined by the corresponding compactification volumes which can be different. This could (at least partially) account for the 
differences between the observed low energy gauge couplings. An immediate objection to this scenario is that this would require some fine tuning of the compactification volumes, and the coupling unification in the MSSM would have to be a complete accident. The first objection might not be too sound as similar ``fine tuning'' can be argued to be required for, say, orbifold models where all three of the gauge subgroups come from the same set of the parallel branes
(and the bifundamental matter arises due to the orbifold projections)\footnote{I would like to thank Cumrun Vafa for pointing this out.}. In such compactifications the gauge couplings of the individual subgroups at the string scale are controlled by VEVs of certain closed twisted sector scalars \cite{DM}, and these VEVs must be fine-tuned to achieve equality of the gauge couplings at the string scale. Here we should point out that these VEVs might simply be zero             
once supersymmetry is broken as the corresponding soft masses (more
precisely, their squares) can be positive at the origin (and this could occur dynamically). However, then we can ask whether in a scenario where different gauge subgroups come from different branes some dynamics can fix the corresponding compactification radii in a fashion consistent with the low energy data. Moreover, one could ask whether this mechanism could also explain the unification of couplings in the MSSM so it does no longer seem to be a complete accident. One idea along these lines was proposed in \cite{BW} where it was pointed out that in F-theory backgrounds with varying dilaton the gauge coupling ``deunification'' could (at least partially) be explained by the fact that the string coupling is different at different branes. It would be interesting to understand this and other possible dynamical mechanisms for solving the gauge coupling unification problem in more detail. 

\acknowledgments

{}I would like to thank Nima Arkani-Hamed, Savas Dimopoulos and Gia Dvali for collaboration at initial stages of this work and valuable discussions. I would also like to thank Pran Nath, Henry Tye, Cumrun Vafa, and especially Tom Taylor for useful discussions. 
This work was supported in part by the grant NSF PHY-96-02074, 
and the DOE 1994 OJI award. I would also like to thank Albert and Ribena Yu for 
financial support.

\appendix
\section{Kaluza-Klein Thresholds}

{}In this appendix we would like to sketch the computation which leads to the expression for the KK thresholds (\ref{thres}).
To begin with, let us consider the ${\cal N}=2$ theory with ${\widetilde \Phi}$ superfields which arises upon a 
compactification of the corresponding six dimensional ${\cal N}=1$ supersymmetric  
gauge theory (living in the world-volume of D5-branes) on $S^1\otimes S^1$. In the following we will be a bit more general and treat it as a compactification of a $p+1$ dimensional theory
(living in the world-volume of Dp-branes) on a product of $(p-3)$ identical circles of radius $R$. The KK spectrum of this theory consists of the states with masses $M_{\bf m}^2={\bf m}^2/R^2$
(${\bf m}=(m_1,\dots,m_{p-3}$), $m_1,\dots,m_{p-3}\in {\bf Z}$) with ${\widetilde \Phi}$ quantum numbers. These states are ${\cal N}=2$ supersymmetric from the four dimensional viewpoint. They contribute into the renormalization of the low energy gauge couplings. Note that perturbatively there are no corrections beyond one loop in this theory, which is due to ${\cal N}=2$ supersymmetry. Moreover, in the D-brane context {\em no other} states contribute to the gauge coupling renormalization. This follows from the fact that in perturbative open string theory only BPS states can renormalize gauge couplings \cite{DL,bac}. In six dimensional ${\cal N}=1$ open string theories the only BPS states are the massless states, whereas all the other states are non-BPS as they come in six dimensional ${\cal N}=2$ (that is, four dimensional ${\cal N}=4$) supermultiplets. The latter, however, do not renormalize gauge couplings. We are, therefore, left only with the contributions of the KK modes (arising upon the compactification on $S^1\otimes S^1$) of the massless modes in six dimensions. Here we outline the calculation of the corresponding KK thresholds.  

{}The computation of the gauge coupling renormalization due to the KK modes in this ${\cal N}=2$ theory is actually a field theoretic computation. In fact, the only place where string theory
becomes relevant for this computation is in the discussion of the ultra-violet (UV) cut-off. The standard Coleman-Weinberg prescription gives the following simple result\cite{TV}:     
\begin{equation}
 \alpha^{-1}_a(\mu)=\alpha^{-1}_a (\Lambda)
 +{{\widetilde b}_a\over 4\pi} \int_{(\xi\Lambda)^{-2}}^{(\xi \mu)^{-2}}
 {dt\over t} \sum_{\bf m} \exp(-\pi t M_{\bf m}^2)~.
\end{equation}
Here $\mu$ and $\Lambda$ are the IR respectively UV cut-offs, and we have parametrized the subtraction scheme dependence by $\xi$. Next, we will identify $\Lambda$ with the string scale
$M_s$, and $\alpha_a (\Lambda)$ with $\alpha$. This way we will obtain the expected logarithmic evolution of the low energy ($\mu\ll 1/R<M_s$) gauge couplings $\alpha_a(\mu)$.
This logarithmic contribution comes from the massless modes with ${\bf m}=0$:
\begin{equation}
 \alpha^{-1}_a(\mu)=\alpha^{-1}
 +{{\widetilde b}_a\over 2\pi} \log\left({M_s\over\mu}\right) +{\widetilde \Delta}_a~.
\end{equation}
The IR {\em finite} threshold corrections ${\widetilde \Delta}_a$ are due to the massive KK modes with ${\bf m}\not=(0,\dots,0)$. The leading contribution (in the regime $(RM_s)^{p-3}\gg1$) to ${\widetilde \Delta}_a$ can be
readily evaluated using the Poisson resummation, and the result is given by \cite{TV}: 
\begin{equation}\label{thresN=2}
 {\widetilde \Delta}_a ={{\widetilde b}_a\over 2\pi} {\xi^{p-3}\over {p-3}} (RM_s)^{p-3}
 -{{\widetilde b}_a\over 2\pi} \log\left(R M_s\right)
 +{\cal O}(1)~.
\end{equation}
Note that the subtraction scheme dependent parameter $\xi$ cannot be determined within these considerations alone\footnote{\label{foot}Here we note that
in \cite{dien} the choice of the subtraction scheme was such that $\xi^{p-3}=\pi^{(p-3)/2}/\Gamma((p-1)/2)$.}. However, in a given theory $\xi$ affects the unification scale $M_s$ (for given values of the low energy gauge couplings). It could, therefore, be determined by fixing $M_s$ via some other low energy quantity (such as the Newton's constant $G_N$). 

{}It is now straightforward to deduce the massive KK threshold corrections (\ref{thres}) in the model of section II. In fact, in this case we have $\Delta_a={\widetilde \Delta}_a/2$, where ${\widetilde \Delta}_a$ are the massive KK threshold corrections (\ref{thresN=2}) in the parent ${\cal N}=2$ theory. This simply follows from the fact that the number of the massive KK states in the ${\cal N}=1$ theory is two times smaller than in the parent ${\cal N}=2$ theory due to the ${\bf Z}_2$ orbifold action. Thus, we arrive at (\ref{thres}) with
\begin{equation}
 \eta_p={1\over 2} {\xi^{p-3}\over {p-3}}~.
\end{equation}
The fact that the coefficient $\eta_p$ is two times smaller
than in the parent ${\cal N}=2$ theory is not surprising: the ${\bf Z}_2$ orbifold
reduces the volume of the $(p-3)$ compact
coordinates inside of the D$p$-brane world-volume 
by a factor of two - the compactification space is $(S^1)^{p-3}/{\bf Z}_2$. That is,
\begin{eqnarray}
 V_{p-3}= {1\over 2} (2\pi R)^{p-3}~.
\end{eqnarray}
This is consistent with the fact that the leading power-like contribution to the massive KK thresholds should scale with the compactification volume $V_{p-3}$ in the same way as the inverse tree-level gauge coupling $\alpha$.


\begin{references} 

\bibitem{gut}
S. Dimopoulos and H. Georgi, Nucl. Phys. {\bf B150} (1981) 193;\\ 
W.J. Marciano and G. Senjanovi{\'c}, Phys. Rev. {\bf D25} (1982) 3092;\\
C. Giunti, C.W. Kim and U.W. Lee, Mod. Phys. Lett. {\bf A6} (1991) 1745; \\
J. Ellis, S. Kelley and D.V. Nanopoulos, Phys Lett. {\bf B249} (1990) 441;\\
U. Amaldi, W. de Boer and H. Furstenau, Phys. Lett. {\bf B260} (1991) 447;\\
P. Langacker and M. Luo, Phys. Rev. {\bf D44} (1991) 817.\\
For a review of the string perspective, see, {\em e.g.},\\
K.R. Dienes, Phys. Rept. {\bf 287} (1997) 447.

\bibitem{BW} For a recent discussion, see, {\em e.g.},\\
Z. Kakushadze and S.-H.H. Tye, ``Brane World'', hep-th/9809147, and references therein.

\bibitem{witt} E. Witten, Nucl. Phys. {\bf B471} (1996) 135.

\bibitem{lyk} J. Lykken, Phys. Rev. {\bf D54} (1996) 3693.

\bibitem{TeV} N. Arkani-Hamed, S. Dimopoulos and G. Dvali, 
Phys. Lett. {\bf B429} (1998) 263.

\bibitem{anto} I. Antoniadis, N. Arkani-Hamed, S. Dimopoulos and G. Dvali,
Phys. Lett. {\bf B436} (1998) 257.

\bibitem{ST} G. Shiu and S.-H.H. Tye, Phys. Rev. {\bf D58} (1998) 106007.

\bibitem{TeVphen} N. Arkani-Hamed, S. Dimopoulos and G. Dvali, 
hep-ph/9807344.

\bibitem{dien} K.R. Dienes, E. Dudas and T. Gherghetta, Phys. Lett. {\bf B436} (1998) 55;
hep-ph/9806292; hep-ph/9807522.

\bibitem{TV} T.R. Taylor and G. Veneziano, Phys. Lett. {\bf B212} (1988) 147.

\bibitem{related} 
A. Pomarol and M. Quir{\'o}s, hep-ph/9806263;\\
C.P. Bachas, hep-ph/9807415;\\
P.C. Argyres, S. Dimopoulos and J. March-Russell, hep-th/9808138;\\
N. Arkani-Hamed, S. Dimopoulos and J. March-Russell, hep-th/9809124;\\
D. Ghilencea and G.G. Ross, hep-ph/9809217;\\
K.R. Dienes, E. Dudas, T. Gherghetta and A. Riotto, hep-ph/9809406;\\
S. Abel and S. King, hep-ph/9809467;\\
D. Lyth, hep-ph/9810320;\\
I. Antoniadis, S. Dimopoulos, A. Pomarol and M. Quir{\'o}s, hep-ph/9810410;\\
G.F. Giudice, R. Rattazzi and J.D. Wells, hep-ph/9811291;\\
S. Nussinov and R. Shrock, hep-ph/9811323;\\
E.A. Mirabelli, M. Perelstein and M.E. Peskin, hep-ph/9811337;\\
T. Han, J.D. Lykken and R.-J. Zhang, hep-ph/9811350;\\ 
N. Kaloper and A. Linde, hep-th/9811141;\\
J.L. Hewett, hep-ph/9811356.

\bibitem{other} 
H. Hatanaka, T. Inami and C.S. Lim, hep-th/9805067;\\
R. Sundrum, hep-ph/9805471, hep-ph/9807348;\\
K. Benakli, hep-ph/9809582;\\
L. Randall and R. Sundrum, hep-th/9810155;\\
K. Benakli and S. Davidson, hep-ph/9810280;\\
G.F. de T{\'e}ramond, hep-ph/9810436;\\
C.P Burgess, L.E. Ib{\'a}{\~n}ez and F. Quevedo, hep-ph/9810535;\\
M. Maggiore and A. Riotto, hep-th/9811089;\\
M. Drees, O.J.P. {\'E}boli and J.K. Mizukoshi, hep-ph/9811343.

\bibitem{quiros} 
I. Antoniadis, Phys. Lett. {\bf B246} (1990) 377;\\
I. Antoniadis, C. Mu{\~n}oz and M. Quir{\'o}s, Nucl. Phys. {\bf B397}
(1993) 515;\\
I. Antoniadis, K. Benakli and M. Quir{\'o}s, Phys. Lett. {\bf B331} (1994)
313;\\
I. Antoniadis, S. Dimopoulos and G. Dvali, Nucl. Phys. {\bf B516} (1998) 70.

\bibitem{polchi} J. Polchinski, Phys. Rev. Lett. {\bf 75} (1995) 4724.

\bibitem{kap} V. Kaplunovsky, Nucl. Phys. {\bf B307} (1988) 145; (E) Nucl. Phys. {\bf B382} (1992) 346.

\bibitem{pol} J. Polchinski, ``TASI Lectures on D-Branes'', hep-th/9611050.

\bibitem{bac} C. Bachas and C. Fabre, Nucl. Phys. {\bf B476} (1996) 418.

\bibitem{thooft} G. 't Hooft, Nucl. Phys. {\bf B72} (1974) 461.

\bibitem{SV} M.A. Shifman and A.I. Vainshtein, Nucl. Phys. {\bf B277} (1986) 456.\\
For a recent discussion, see, {\em e.g.},\\
N. Arkani-Hamed and H. Murayama, Phys. Rev. {\bf D57} (1998) 6638; hep-th/9707133.

\bibitem{Jones} P.M. Ferreira, I. Jack, D.R.T. Jones and C.G. North, Nucl. Phys. {\bf B504} (1997) 108.

\bibitem{Ross} G. Amelino-Camelia, D. Ghilencea and G.G. Ross, Nucl. Phys. {\bf B528} 
(1998) 35.

\bibitem{flavor} N. Arkani-Hamed and S. Dimopoulos, hep-ph/9811353;\\
Z. Berezhiani and G. Dvali, hep-ph/9811378.

\bibitem{typeI}
M. Berkooz and R.G. Leigh, Nucl. Phys. {\bf B483} (1997) 187;\\
C. Angelantonj, M. Bianchi, G. Pradisi, A. Sagnotti and 
Ya.S. Stanev, Phys. Lett. {\bf B385} (1996) 96;\\
Z. Kakushadze, Nucl. Phys. {\bf B512} (1998) 221;\\ 
Z. Kakushadze and G. Shiu, Phys. Rev. {\bf D56} (1997) 3686; Nucl. Phys. {\bf B520} (1998) 75;\\
G. Zwart, Nucl. Phys. {\bf B526} (1998) 378;\\
G. Aldazabal, A. Font, L.E. Ib{\'a}{\~n}ez and G. Violero, hep-th/9804026;\\
R. Blumenhagen and A. Wisskirchen, Phys. Lett. {\bf B438} (1998) 52;\\
J.D. Lykken, E. Poppitz and S. Trivedi, hep-th/9806080;\\
L.E. Ib{\'a}{\~n}ez, R. Rabadan and A.M. Uranga, hep-th/9808139.

\bibitem{KST} Z. Kakushadze, G. Shiu and S.-H.H. Tye, Nucl. Phys. {\bf B533} (1998) 25.

\bibitem{3gen} Z. Kakushadze, Phys. Lett. {\bf B434} (1998) 269; Nucl. Phys. {\bf B535} (1998) 311; Phys. Rev. {\bf D58} (1998) 101901;\\
Z. Kakushadze and S.-H.H. Tye, Phys. Rev. {\bf D58} (1998) 126001.

\bibitem{Voisin} C. Voisin, in: Journe{\'e}s de G{\'e}om{\'e}trie 
Alg{\'e}brique d'Orsay, eds. A. Beauville {\em et al.},
Ast{\'e}risque, vol. 218 (Soc. Math. France, 1993) 273-323;\\
C. Borcea, in: Mirror Manifolds II, eds. B.R. Greene and
S.-T. Yau (International Press and AMS, 1997) 717-743;\\
For a physicist's discussion, see, {\em e.g.},\\
P.S. Aspinwall, Nucl. Phys. {\bf B460} (1996) 57;\\
D.R. Morrison and C. Vafa, Nucl. Phys. {\bf B473} (1996) 74; 
Nucl. Phys. {\bf B476} (1996) 437.

\bibitem{aspin} P.S. Aspinwall, Phys. Lett. {\bf B357} (1995) 329.

\bibitem{sen} A. Sen, Phys. Rev. {\bf D55} (1997) 7345.

\bibitem{vafa} C. Vafa, Nucl. Phys. {\bf B469} (1996) 403.  

\bibitem{INQ} L.E. Ib{\'a}{\~n}ez, H.P. Nilles and F. Quevedo, Phys. Lett. {\bf B187} (1987) 25;\\
Z. Kakushadze and S.-H.H. Tye, hep-th/9512156; Phys. Rev. {\bf D54} (1996) 7520. 

\bibitem{KaSi}
S. Kachru and E. Silverstein, Phys. Rev. Lett. {\bf 80} (1998) 4855;\\
A. Lawrence, N. Nekrasov and C. Vafa, Nucl. Phys. {\bf B533} (1998) 199;\\
M. Bershadsky, Z. Kakushadze and C. Vafa, Nucl. Phys. {\bf B523} (1998) 59.

\bibitem{BaDi} M. Dine and N. Seiberg, Phys. Rev. Lett. {\bf 55} (1985) 366;
Phys. Lett. {\bf B162} (1985) 299;\\
V.S. Kaplunovsky, Phys. Rev. Lett. {\bf 55} (1985) 1036;\\
T. Banks and M. Dine, Phys. Rev. {\bf D50} (1994) 7454.

\bibitem{kras} See, {\em e.g.},\\
N.V. Krasnikov, Phys. Lett. {\bf B193} (1987) 37; \\
L. Dixon, V. Kaplunovsky, J. Louis and M. Peskin, SLAC-PUB-5229 (1990); \\
J.A. Casas, Z. Lalak, C. Mu{\~n}oz and G.G. Ross, Nucl. Phys. {\bf B347} (1990)
243; \\
T.R. Taylor, Phys. Lett. {\bf B252} (1990) 59;\\
V.S. Kaplunovsky and J. Louis, Phys. Lett. {\bf B417} (1998) 45;\\
G. Dvali and Z. Kakushadze, Phys. Lett. {\bf B417} (1998) 50;\\
M. Klein and J. Louis, hep-th/9803143.

\bibitem{Alda} G. Aldazabal, A. Font, L.E. Ib{\'a}{\~n}ez, A.M. Uranga and G. Violero, Nucl. Phys. {\bf B519} (1998) 239. 

\bibitem{HW} P. Ho{\u r}ava and E. Witten, Nucl. Phys. {\bf B460} (1996) 506; Nucl. Phys. {\bf B475} (1996) 94.

\bibitem{ovrut} R. Donagi, A. Lukas, B.A. Ovrut and D. Waldram, hep-th/9811168.

\bibitem{DM} M.R. Douglas and G. Moore, hep-th/9603167.

\bibitem{DL} See, {\em e.g.},\\
W. Lerche, Nucl. Phys. {\bf B308} (1988) 102;\\ 
V. Kaplunovsky and J. Louis, Nucl. Phys. {\bf B444} (1995) 191;\\
M. Bershadsky, S. Cecotti, H. Ooguri and C. Vafa, Commun. Math. Phys. {\bf165} (1994) 311;\\
J.A. Harvey and G. Moore, Nucl. Phys. {\bf B463} (1996) 315;\\
M.R. Douglas and M. Li, hep-th/9604041.

\end{references}
\end{document}